\documentclass[11pt,onecolumn,floatfix,aps,nofootinbib]{revtex4-1}
\usepackage[utf8]{inputenc}
\usepackage{array}
\usepackage{amsmath}
\usepackage{slashed}
\usepackage{amsfonts}
\usepackage{amssymb} 
\usepackage{wasysym}
\usepackage[all,cmtip]{xy}
\usepackage[english]{babel}
\usepackage{bbm}
\usepackage{caption}
\usepackage{verbatim,upgreek}
\usepackage{array}
\usepackage{multirow}
\usepackage{amsfonts}
\usepackage{slashed} 
\usepackage{hyperref}
\usepackage{leading}
\usepackage[dvipsnames]{xcolor}
\hypersetup{colorlinks=true,linkbordercolor=Blue,linkcolor=Blue, citecolor=Blue}
\usepackage{subcaption}
\usepackage{graphicx}
\usepackage{subeqnarray}
\usepackage{setspace}
\newcommand{\gdualn}[1]{\overset{\:{}^{{}^{\boldsymbol{\neg}}}}{\smash[t]{#1}}} 
\usepackage{indentfirst} 
\usepackage{gensymb}
\newsavebox\myboxA
\newsavebox\myboxB
\newlength\mylenA

\begin{document}

\title{Sensitivity of accelerator-based neutrino experiments to neutrino-dark matter interaction: elko as a viable dark matter candidate.}

\author{C. A. Moura$^{1}$} \email{celio.moura@ufabc.edu.br}
\author{F. Kamiya$^{1}$} \email{felipe.kamiya@ufabc.edu.br}
\author{L. Duarte$^{2}$} \email{l.duarte@unesp.br}
\author{M. Dias$^{3}$} \email{marco.dias@unifesp.br}
\author{J. M. Hoff da Silva$^{2}$} \email{julio.hoff@unesp.br}

\affiliation{$^{1}$Universidade Federal do ABC (UFABC), Santo Andr\'e, 09210-580, Brazil}
\affiliation{$^{2}$Departamento de F\'isica, Universidade
Estadual Paulista, UNESP, Av. Dr. Ariberto Pereira da Cunha, 333, Guaratinguet\'a, SP,
Brazil}
\affiliation{$^{3}$Departamento de F\'isica, Universidade Federal de S\~ao Paulo (UNIFESP), Diadema-SP-Brazil}

\begin{abstract}
We analyze the potential of accelerator-based neutrino experiments, such as the Deep Underground Neutrino Experiment, to constrain a five dimensional operator mediating a beyond standard model interaction between a mass dimension one fermion field, labeled elko, and neutrino. We identify the parameter space where elko is a viable dark matter candidate.
\end{abstract}

\maketitle

\section{Introduction}

In this paper we investigate possible couplings between a fermionic field endowed with canonical mass dimension one and neutrinos under the auspices of accelerator-based neutrino experiments. We use as a case study the Deep Underground Neutrino Experiment (DUNE)~\cite{DUNE:2020lwj,DUNE:2020ypp,DUNE:2020mra,DUNE:2020txw,DUNE:2021tad}.

There are several current candidates in high energy physics to describe dark matter. In a broad brush, many approaches make use of standard fields, but with effective or modified interactions, while some of them are endowed with ad hoc impositions in order to satisfy the necessary cosmological constraints \cite{masi}. Other approaches result as consequence of more complete and comprehensive theoretical scopes \cite{RD}. Whatever the model or theory at hand, finding the relation between a given dark matter candidate and a physical observable is always a nontrivial task. In this regard, the study of theoretical dark matter proposals along with current experiments is certainly a welcome branch of research.

The mass dimension one fermionic field is based upon the so-called elkos, eigenspinors of the charge conjugation operator with dynamics not dictated by the Dirac first order equation. Its neutrality under usual gauge interactions provides a dark matter candidate \cite{ahl}. In trying to envisage possible accelerator signals due to allowed interactions for this specific field, we point out Refs. \cite{vel,nnv} for the early days of the theory, and Refs. in \cite{colet,Alves:2017joy} searching for outcomes via Higgs couplings and monophoton events. 

DUNE belongs to the next generation of long baseline neutrino oscillation experiments and will use two facilities of the United States Department of Energy. A neutrino beam will be produced at the Fermi National Accelerator Laboratory (Fermilab), where the ND will be installed~\cite{DUNE:2021tad}.
The ND is functionally similar to the FD, thus reducing the systematic uncertainties of the experiment. 
It is expected to {\it i)} characterize the beam close to the source in terms of flavor, energy range, and flux of neutrinos; {\it ii)} measure the $\nu$ cross section with the detector material; and {\it iii)} crosscheck the beam models.

The neutrino beam crosses the Earth's crust reaching the Far Detector (FD)~\cite{DUNE:2020lwj}, which uses Liquid Argon Time Projection Chamber (LArTPC) technology and will be installed at the Long Baseline Neutrino Facility (LBNF)~\cite{DUNE:2020mra} in Sanford.
The LBNF will have four LArTPC modules, each with 17 kton of liquid argon of which 10 kton will be the active mass.
The neutrino beam generated at Fermilab is expected to have a power of $1.2$\,MW with an upgrade in the future reaching up to $2.4$\, MW. This initial power for the beam is related to a primary proton with energy of $120$\, GeV. The beam operating at 1.2 MW for one year corresponds to $10^{21}$ protons with 120 GeV colliding with the target within this time period. This process will produce a neutrino beam with most of the produced neutrinos with energies between 1 and 3 GeV.

Because of the high intensity beam and the capability of realizing high precision measurements, DUNE has also a great potential to investigate beyond standard model physics~\cite{Arguelles:2019xgp,DUNE:2020fgq}. Based in this characteristic, we constrain for the first time neutrino-elko interaction in a wide range of masses, from ultralight candidates with $\approx 10^{-30}$~eV to candidates as heavy as TeV scale. We show the constraints for this interaction considering different amounts of elko as a dark matter component.

This paper is organized as follows: for completeness, in the next section we depict some of the general aspects of the mass dimension one fermion formulation. In Section III the neutrino-elko scattering cross section is obtained. Section IV is devoted to the analysis methodology used in this work. The results are presented in Section V. In the final Section we conclude. 

\section{Basic aspects of mass dimension one fermionic field}\label{sec:bas}

The standard protocol setting up a relativistic bispinor makes use of elements of the whole representation space, $(1/2,0)\oplus (0,1/2)$, relating both parts of it, the left and right sides, by means of the parity operator $P$. Elements of the mentioned representation space, whose interchange between the sectors $(1/2,0)$ and $(0,1/2)$ are not performed by $P$, may not obey the Dirac equation \cite{Spera}. This is the first ingredient to depart from the usual fermionic theory. Along with eigenspinor relations under acting of the charge conjugation operator, $C$, the mentioned unusual behavior under $P$ gives rise to elkos, eigenspinors of $C$ obeying the relativistic Klein-Gordon dynamical equation, but not the Dirac equation. 

The spinors resulting from the considerations just previously listed are, then, used as expansion coefficients of a quantum field which obeys fermionic (anti)commutation rules and present a Klein-Gordon propagator. Moreover, the locality structure of the field and Lorentz invariance of the theory is reached by means of a dual field different from the usual one, but encompassed in a judicious theory \cite{ahl, aaca} and in agreement with Clifford algebra possibilities \cite{out}. The way this field circumvent the Weinberg no-go theorem is also inputted to its different dual~\cite{ahlepl}. 

We depict here some aspects of the mass dimension one field, without going much into details on its formal aspects and/or subtleties about the new dual. All these aspects may be found in Ref. \cite{ahl}. Here and there, along the text, some properties as spin sums (eventually computed with the Dirac dual, for particular input in some calculation) are used. In those cases, proper references are cited. 

Elkos are formulated to be eigenspinors of $C$, as $C\lambda^{S/A}=\pm \lambda^{S/A}$. The plus (minus) sign reserved to the label $S$ ($A$). $S$ and $A$ stand for self-conjugate and anti-self-conjugate, respectively.
Each spinor inside the bispinors $\lambda^{S/A}$ is obtained in the helicity basis, i.e., they are eigenspinors (with eigenvalues $+1$ and $-1$) of $(\sigma\cdot \hat{p})/p$, resulting in a set of four spinors $\{\lambda^S_{+}, \lambda^S_{-},\lambda^A_{+}, \lambda^A_{-} \}$ whose dynamics is given by the Klein-Gordon equation. The formalization of these spinors can be found in Refs. \cite{ahl,aaca}.

Using the aforementioned spinors as expansion coefficients, the quantum field is defined as 
\begin{equation}
\eta(x)=\int\frac{d^3{\bf p}}{(2\pi)^3}\frac{1}{\sqrt{2m_\lambda E({\bf p})}}\sum_{\alpha}\Bigg(a_\alpha({\bf p})\lambda^S_\alpha({\bf p})e^{-ip\cdot x}+b^\dagger_\alpha({\bf p})\lambda^A_\alpha({\bf p})e^{ip\cdot x}\Bigg), \label{1}
\end{equation} while its adjoint is given by 
\begin{equation}
\gdualn{\eta}=\int\frac{d^3{\bf p}}{(2\pi)^3}\frac{1}{\sqrt{2m_\lambda E({\bf p})}}\sum_{\alpha}\Bigg(a_\alpha({\bf p})\gdualn{\lambda}^S_\alpha({\bf p})e^{ip\cdot x}+b^\dagger_\alpha({\bf p})\gdualn{\lambda}^A_\alpha({\bf p})e^{-ip\cdot x}\Bigg). \label{2}
\end{equation} 
As usual, $p$ stands for the four-momentum, $E$ for the energy, $m_\lambda$ for the mass, and $x$ for the space-time coordinates. The fields obey the Fermi-Dirac statistics \cite{ahl} and therefore the annihilation and creation operators for particles and antiparticles satisfy
$\{a_\alpha ({\bf p}),a^\dagger_{\alpha'} ({\bf q})\}=\{b_\alpha ({\bf p}),b^\dagger_{\alpha'} ({\bf q})\}=(2\pi)^3\delta^{(3)}({\bf p}-{\bf q})\delta_{\alpha,\alpha'}$, with all other anticommutator combinations vanishing. 

There are important remarks concerning elko spin sums which we shall address here, since it will be import in computing the cross section. For details see \cite{PR}. Using the standard Dirac dual for elko, the scalar bilinear is vanishing and, hence, in order to avoid this behavior, it is defined an intermediate dual given by
\begin{equation}
    \tilde{\lambda}^{S/A}_\pm (p^\mu)=\mp i(\lambda^{S/A}_\mp (p^\mu))^\dagger\gamma^0,\label{ref1}
\end{equation} by means of which the following orthonormality relations for arbitrary momentum hold
\begin{equation}
    \tilde{\lambda}^{S}_\alpha\lambda^S_{\beta}=-\tilde{\lambda}^A_\alpha\lambda^A_\beta=2m_{\lambda} \delta_{\alpha\beta},\label{ref2}
\end{equation} while any other product is null. The spin sums computed with this dual are given by
\begin{eqnarray}
\sum\limits_{\alpha}{\lambda}_{\alpha}^{S}({p^{\mu}})\tilde{\lambda}_{\alpha}^{S}({p^{\mu}})=m_\lambda(I+\mathcal{G}(\varphi)),\nonumber\\
\sum\limits_{\alpha}{\lambda}_{\alpha}^{A}({p^{\mu}})\tilde{\lambda}_{\alpha}^{A}({p^{\mu}})=-m_\lambda(I-\mathcal{G}(\varphi)), \label{ref3}
\end{eqnarray} where the matrix $\mathcal{G}(\varphi)$ has non-null elements only in the secondary diagonal \cite{Ahluwalia:2004ab}
\begin{eqnarray}
\mathcal{G}(\varphi)=
\begin{pmatrix}
0 & 0 & 0 & -ie^{-i\varphi}\\
0 & 0 & ie^{i\varphi} & 0 \\
0 & -ie^{-i\varphi} & 0 & 0 \\
ie^{i\varphi} & 0 & 0 & 0
\end{pmatrix}.
\end{eqnarray} The spin sums (\ref{ref3}) are clearly not Lorentz invariant. A procedure was developed by the introduction of operators serving, after a procedure so-called $\tau-$deformation, as appropriate inverse of the Lorentz broken factors in (\ref{ref3}) (see \cite{PR} for a mathematical and physical account on this procedure). This lead to a new dual preserving the orthonormal conditions in terms of which the spin sums read
\begin{eqnarray}
\sum_{\alpha} \lambda^S_\alpha(p^\mu)\gdualn{\lambda}^S_\alpha(p^\mu)=2m_\lambda\mathbb{I}\,,\nonumber\\
\sum_{\alpha} \lambda^A_\alpha(p^\mu)\gdualn{\lambda}^A_\alpha(p^\mu)=-2m_\lambda\mathbb{I}\,. \label{3}
\end{eqnarray}

The standard correlator, with the above ingredients, reads 
\begin{equation}
\langle 0 | \mathcal{T} \eta(x)\gdualn{\eta}(x') | 0 \rangle = i\int \frac{d^4p}{(2\pi)^4} \frac{\mathbb{I}}{p^2-m_\lambda^2}e^{ip\cdot (x-x')},\label{4}
\end{equation} from which the Klein-Gordon propagator and, hence, the canonical dimensionality can be read. In Eq. (\ref{4}), $\mathcal{T}$ stands for the usual time ordering operator. 

\section{Neutrino-Elko Scattering Cross Section}

The general form of a scattering process is governed by the dynamics encoded in the transition matrix elements
and the phase space available in the interaction.

The Lagrangian describing the new interaction is:
\begin{eqnarray}\label{lagrangiana}
\mathcal{L}=\frac{\xi}{\Lambda}~
\overline{\nu}_{F} \nu_{F'}
\stackrel{\neg}{\lambda}
_{\chi}{\lambda}_{\chi'}\,,
\end{eqnarray}
where $\xi$ is the coupling strength, $\Lambda$ is the energy scale of the theory and has mass dimension one, $\nu$ stands for a neutrino and ${\lambda}$ represents 
an elko. $F$ and $F'$ are lepton
flavors, $e,\mu$, and $\tau$. $\chi$ and $\chi'$ are the type of eigenspinor under charge transformation.
When we decompose these fields in their normal modes, $\chi^{(')}=S$ corresponds to particle states and $\chi^{(')}=A$ to antiparticle states.
Because we are working with particle states only, from now on we suppress the index $S$ used in Sec.~\ref{sec:bas}.

\begin{figure}[h!]
\includegraphics[scale=.2]{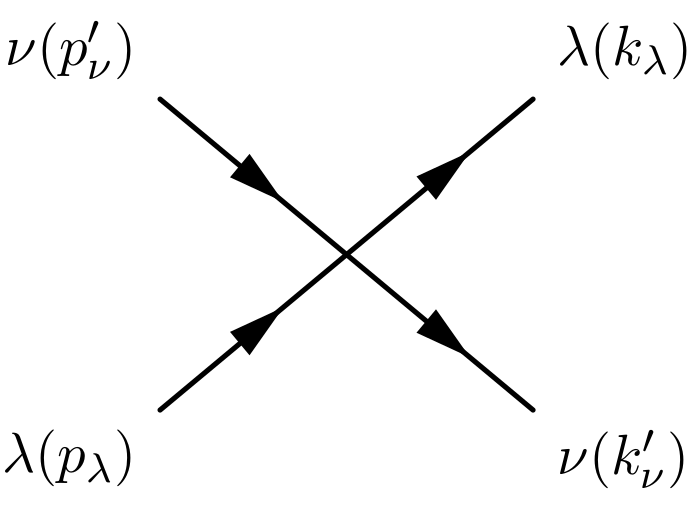}
\vspace{4mm}
\caption{Feynman diagram for neutrino-elko scattering. See text for details.}
\label{fd}
\end{figure}

For the Lagrangian described in Eq. \eqref{lagrangiana}, we are interested in the process displayed in Fig.~\ref{fd}.
\vspace{4mm}
In this case, a neutrino with four-momentum $p'_{\nu}$ (aligned along the $z-$direction) scatters off an elko with four-momentum $p_{\lambda}$, which is at rest in the lab frame. This produces an outgoing neutrino with four-momentum $k'_{\nu}$ and a scattered elko with four-momentum $k_\lambda$. 
The neutrino mass scale is much lower than the electron and the elko masses, and also comparing to a typical neutrino beam energy. Therefore, we neglect the neutrino mass in this analysis. 
In the lab frame, the components of these quantities can be written as:
\begin{eqnarray}\label{4momentum}
&&p'_{\nu}=(E_{\nu_i},{\bf p'})\,;\nonumber \\
&&k'_{\nu}=(E_{\nu_f}, {\bf k'})\, \nonumber;\\
&&p_{\lambda}=(m_\lambda,{\bf 0})\,\nonumber;\\
&&k_\lambda=(T+m_\lambda, {\bf k_{\lambda}})\,,
\end{eqnarray}
where $T$ is the outgoing elko kinetic energy, $m_\lambda$ is the mass of the elko, $E_{\nu_i}$ and $E_{\nu_f}$ represent the neutrino energy before and after the interaction, respectively. The boldface letters are the respective three-momentum.

We are interested in the cross section, 
from which we can compute the new energy spectrum at the detector.
In the case of two-body collisions between an incoming neutrino and a (stationary) target elko, through a general nonstandard interaction, the differential cross section is given by the expression \cite{Bischer:2018zcz}, in natural units,
\begin{eqnarray}\label{xsec}
\frac{d\sigma}{dT}=\frac{|\bar{\mathcal{M}}|^2}{32 \pi E_{\nu_i}^2 m_\lambda}\,,
\end{eqnarray}
where $\bar{\mathcal{M}}$ is the unpolarized scattering amplitude, i.e., the sum over all helicity states of $|\mathcal{M}|^2$ with $\mathcal{M}$ given by
\begin{eqnarray}\label{M}
 \mathcal{M}= \frac{\xi}{m_\lambda\Lambda}(\tilde{\lambda}_{\alpha'}(k_\lambda) \, \lambda_{\alpha}(p_{\lambda}))\,(\bar{u}_{\beta}({k'_{\nu}})\,u_{\beta'}({ p'_{\nu}})),
\end{eqnarray}
where $u(p'_{\nu})$ and $\bar{u}(k'_{\nu})$ are the Dirac spinors representing the incoming and outgoing neutrino, respectively. Greek indices, $\alpha, \alpha', \beta$, and $\beta'$, correspond to helicities. In order to ensure unitarity in the process, we shall make use of the adjoint defined in Ref. \cite{PR}. Here we just pinpoint the necessary main steps to our purposes. Calling $I$ a generic label denoting self (S) and anti-self (A) conjugated spinors we may recast (\ref{ref1}) as $\tilde{\lambda}_\alpha^I=(\lambda_\alpha^I)^\ddagger\gamma^0$, where $(\lambda^I_\alpha)^{\ddagger}\equiv-i\alpha(\lambda^I_{-\alpha})^\dagger$. From Eq. (\ref{M}) by calling $\mathcal{M}\propto\mathcal{M}_{elko}\mathcal{M}_{usual}$ we have $\mathcal{M}^\ddagger\propto\mathcal{M}^\ddagger_{usual}\mathcal{M}^\ddagger_{elko}$. It turns out, however, that $\mathcal{M}^\ddagger_{usual}=\mathcal{M}^\dagger_{usual}$ and therefore one is able to write the unpolarized scaterring amplitude by 
\begin{eqnarray}
&&|\bar{\mathcal{M}}|^2=\frac{1}{2}\frac{\xi^2}{m_{\lambda}^2\Lambda^2}\text{Tr}\Big(\sum\limits_{\beta'}\frac{(1-\gamma^5)}{2}{u}_{\beta'}({p'_{\nu}})\bar{u}_{\beta'}({p'_{\nu}})\sum\limits_{\beta}{u}_{\beta}({k'_{\nu}})\bar{u}_{\beta}({k'_{\nu}})\Big) \nonumber \\
 &&\times\text{Tr}\Big(\sum\limits_{\alpha}{\lambda}_{\alpha}(p_{\lambda})\tilde{\lambda}_{\alpha}(p_{\lambda}) \sum\limits_{\alpha'}{\lambda}_{\alpha'}({k_{\lambda}})\tilde{\lambda}_{\alpha'}({ k_{\lambda}})\Big)\,.
\end{eqnarray} Now, performing the generalization to the $\gdualn{\lambda}^S$ and taking the $\tau-$deformation \cite{PR}, we are left with
\begin{eqnarray}\label{cc}
|\bar{\mathcal{M}}|^2 \rightarrow \frac{1}{2}\frac{\xi^2}{m_{\lambda}^2\Lambda^2}\text{Tr}\Big(\frac{(1-\gamma^5)}{2}\slashed{p'}(\slashed{k'})\Big)\text{Tr}\Big(\sum\limits_{\alpha}{\lambda}_{\alpha}(p_{\lambda})\gdualn{\lambda}_{\alpha}(p_{\lambda}) \sum\limits_{\alpha'}{\lambda}_{\alpha'}({k_{\lambda}})\gdualn{\lambda}_{\alpha'}({ k_{\lambda}})\Big),
\end{eqnarray} where $\sum\limits_{\beta}{u}_{\beta}({k'_{\nu}})\bar{u}_{\beta}({k'_{\nu}})=\slashed{k'}$ was used in the approximation of negligible neutrino mass when compared to the elko mass. Notice that matter effects, as the Mikheyev-Smirnov-Wolfenstein (MSW) one, are not expected in our estimates. Bearing in mind similar calculations as in the case of standard weak interaction, one would have an MSW-like effect if we had usual gauge charged particles involved (like in a standard current interaction), or we would need a neutral current interaction type with changing in neutrino flavor, which is highly constrained. Hence the interaction potential with elko is diagonal in the neutrino evolution equation, giving just a phase, which disappears when we calculate the probability of flavor oscillation. As a last remark, we stress that for the estimate we are working on here, as there is no anti-self conjugated mass dimension one spinors at the final state, an unimodular phase appearing in the mass dimension one fermions cross section construction is taken equal to one \cite{PR}. Finally, by means of the first equation in (\ref{3}) we obtain
\begin{eqnarray}\label{final}
|\bar{\mathcal{M}}|^2 = 16\frac{\xi^2}{\Lambda^2}(p'_\nu\cdot k'_\nu)\,.
\end{eqnarray} Calling $\theta$ the angle between the scattered and incident neutrino momentum, it is not difficult to see that, by means of momentum conservation in the plane of the reaction, $p'_\nu\cdot k'_\nu \simeq |{\bf k'}|\sqrt{|{\bf k_\lambda}|^2-|{\bf k'}|^2\sin^2\theta}$, so that $\sin\theta$ is bounded from above by $|{\bf k_\lambda}|/|{\bf k'}|$. As it is an elastic (point) interaction, bearing in mind that the neutrino mass is small when compared to the elko mass, there is little energy transferred to the elko and we are, therefore, dealing with a typical forward scattering. As we shall see, however, a wide range of elko mass shall be explored here, but even in the low elko mass case we still have forward scattering, since the ratio of elko and neutrino final moments must be small, otherwise the neutrino falls under the energy resolution for detection. The four-momentum relations defined in Eq. \eqref{4momentum} and Eq. \eqref{final} result in the averaged probability amplitude,
\begin{eqnarray}\label{final1}
|\bar{\mathcal{M}}|^2 = 16\frac{\xi^2}{\Lambda^2}m_{\lambda}T.
\end{eqnarray}
Therefore, the differential cross section, Eq.~\eqref{xsec}, for elko-neutrino scattering $\nu_\mu + \lambda \to \lambda + \nu_\mu$ simplifies to
\begin{equation}
    \dfrac{d\sigma_{\nu-\lambda}}{dT} = \frac{\xi^2}{\Lambda^2}\frac{1}{2\pi E_{\nu_i}^2}T\,.
\end{equation}
After integration, taking into account that $T=E_{\nu_i}-E_{\nu_f}$, the total cross section takes the form,
\begin{equation}
\sigma(E_{\nu_i},E_{\nu_f}) = \frac{\xi^2}{4\pi E_{\nu_i}^2\Lambda^2}(E_{\nu_i}-E_{\nu_f})^2\,.
    \label{crosssection_neutrino_ELKO}
\end{equation}

In the next section we present the methodology used to compute the effect that neutrino-elko interaction along the beam trajectory may bring to the number of events.

\section{Methodology} \label{Simulation_Methodology}

\begin{figure}[h!]
\centering
\includegraphics[scale=0.8]{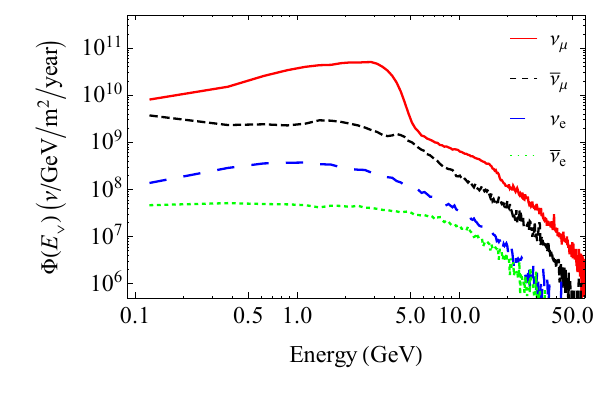}
\caption{Muon neutrino flux and some of the background components as a function of energy in the DUNE FD. Red line stands for muon neutrino, dashed black for muon antineutrino, long dashed blue for electron neutrino, and dotted green for electron antineutrino.}
\label{fluxo_CDR}
\end{figure}

We use the simulation of neutrino flux, $\phi$, at the DUNE FD described and available at~\cite{DUNE:2021cuw}. Fig.~\ref{fluxo_CDR} shows the main components of the neutrino flux as a function of the energy.

The number of events at the DUNE FD is computed using GLoBES~\cite{Huber:2004ka,Huber:2007ji}. We compute the effect of the new neutrino interaction in the neutrino flux, which generates a new energy spectrum at the FD. The interaction cross section in Eq.~\eqref{crosssection_neutrino_ELKO} is used to determine the new flux at the detector according to,
\begin{equation}
\phi(E_{\nu_f}) = \sum_{E_{\nu_i}}
\phi(E_{\nu_i})\sigma(E_{\nu_i},E_{\nu_f})n_\lambda L +  
\left.\left(\phi(E_{\nu_i})-\sum_{E_{\nu}}
\phi(E_{\nu_i})\sigma(E_{\nu_i},E_{\nu})n_\lambda L \right)\right|_{E_{\nu_i}=E_{\nu_f}}\,,
\label{eq:new_flux}
\end{equation}
where $E_\nu$ are the neutrino energies in the flux data, $n_\lambda$ is the average elko number density in the propagation medium, and $L$ is the distance traveled by the neutrino from the source to the detector. The first sum in the right-hand side of Eq~\eqref{eq:new_flux} gives the flux of neutrinos that interacted, changing their original energy to $E_{\nu_f}$. The term in parentheses accounts for the neutrinos that having original energy $E_{\nu_f}$, change their energy to any other value considered in the simulation.

In order to define the density of elkos along the propagation, we assume elkos are at least part of the dark matter content in the Universe~\cite{Ahluwalia:2004sz}, considering a few different elko-dark matter ratios.
We define the relic dark matter as,
\begin{equation}
Y={n}_\lambda/\mathcal{T}^3 \,,    
\end{equation}
where $\mathcal{T}$ is the temperature. If elkos are cold dark matter, then the temperature can be written as,
\begin{equation}
\mathcal{T}=m_\lambda/x_f\,,     
\end{equation}
where $x_f$ is the freeze-out parameter.
The elko density along the baseline, considering it is uniformly distributed, is obtained from the previous two equations~\cite{Dodelson:2003ft},
\begin{equation}
    n_{\lambda} = Y\left(\dfrac{m_\lambda}{x_f}\right)^3\,,
\end{equation}
where $Y$ must be in the range 0.1126 $\leq Y \leq 0.1246$ if all dark matter is composed by elko and $x_f=20$~\cite{ParticleDataGroup:2020ssz}. Values of Y $\leq$ 0.1126 are possible if elko is a subdominant component of the cosmological dark matter, allowing the coexistence of other candidates~\cite{Aghanim:2018eyx}.

With these considerations, we can compute the DUNE FD sensitivity to neutrino-elko interactions using a minimum square method. In this method one defines a $\chi^2$ function~\cite{Huber:2002mx, Tanabashi:2018oca},
\begin{equation}
    \chi^2 = \sum_{i} 2\left( \left( N^s_i - N^\lambda_i \right) + N^\lambda_i \ln \dfrac{N^\lambda_i}{N^s_i} \right) \,,
    \label{chi_square}
\end{equation}
where $N_i^{s}$ is the standard number of events and $N_i^{\lambda}$ is the number of events considering neutrino-elko interaction. The summation is over all the energy bins. We find the minimum of the $\chi^2$ as a function of the elko mass, $m_\lambda$, coupling strength, $\xi/\Lambda$, and density of elkos, $Y$.
We then evaluate the DUNE FD sensitivities.

\section{Results and discussions}
The sensitivity is computed for two free parameters at a time. The confidence level (CL) is then defined as 90\% for $\Delta\chi^2=\chi^2-\chi^2_m=4.61$, where $\chi^2_m$ is the minimum value obtained from Eq.~\eqref{chi_square} given a set of free parameters.
We analyze the DUNE sensitivity in the two dimension parameter space covering $\xi/\Lambda$ and $m_\lambda$.

\begin{figure}[htb]
\centering
\hspace{-0.0cm}{
\includegraphics[scale=0.75]{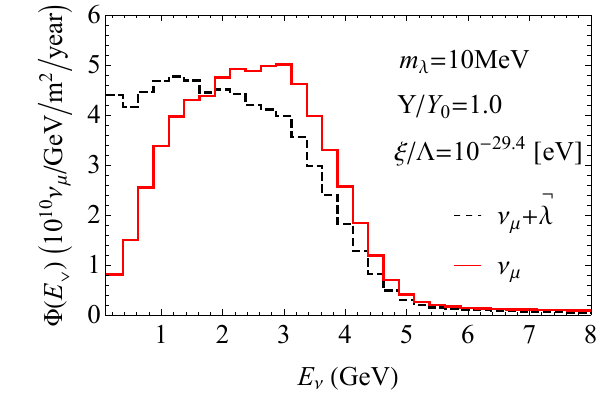}
}
\hspace{-0.0cm}{
\includegraphics[scale=0.75]{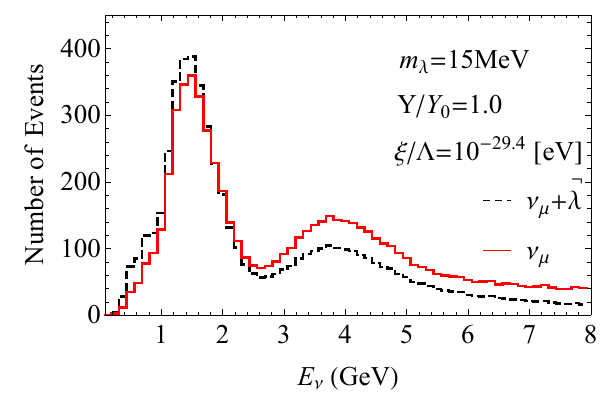}
}
\caption{{\bf Left:} Neutrino flux at the FD. {\bf Right:} Number of events at the FD. In both plots, the red curves represent standard interaction only and the black-dashed curves consider neutrino-elko interaction along the propagation.}
\label{Energy_Spectrum}
\end{figure}

Fig.~\ref{Energy_Spectrum} shows the neutrino flux at the FD on the left plot and the number of events on the right plot. On both plots the red curve represents the standard calculation and the black-dashed represents the calculation considering neutrino interacting with elko along the propagation. As an example, we chose the parameters: $m_{\lambda}=10$~MeV, $\xi/\Lambda=10^{-29.4}$, and $Y=Y_0$, where $Y_0=0.1186$ is the dark matter relic density \cite{Aghanim:2018eyx}. 

Because of the energy loss due to the new interaction, there is an increase of the flux and the number of events at lower energies, while there is a decrease at higher energies.

\begin{figure}[htb]
    \centering
    \includegraphics[scale=0.50]{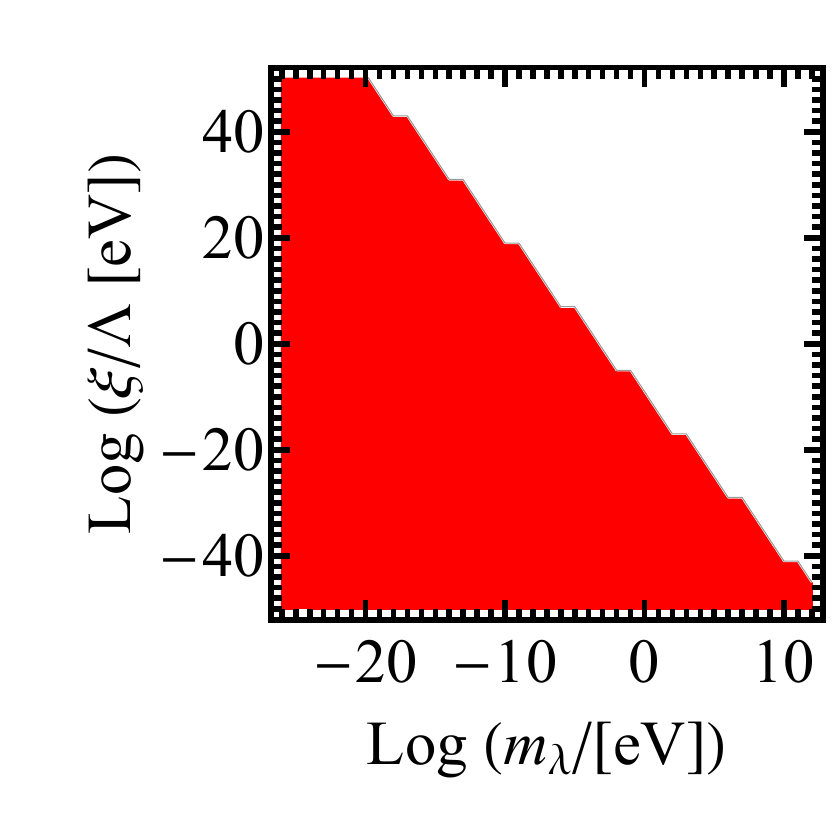}
    \includegraphics[scale=0.50]{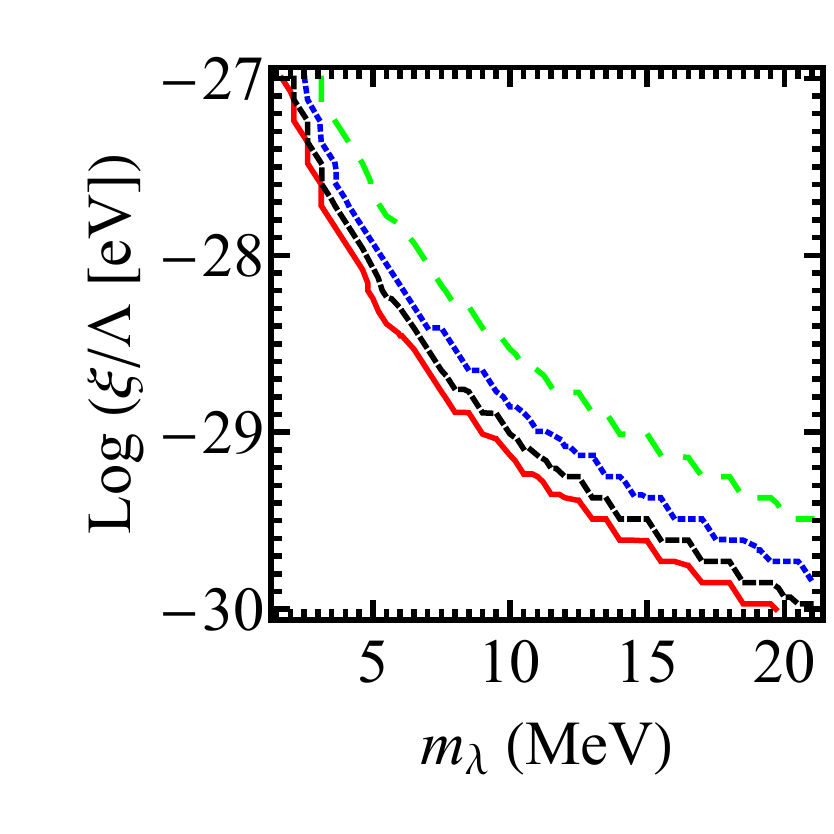}
    \caption{Space of parameters allowed at 90\% CL. {\bf Left:} $\log(\xi/\Lambda)$, where $\xi$ has no dimension and $\Lambda$ has units of eV, as a function of $\log m_\lambda$, where $m_\lambda$ is given in units of eV. The allowed region of the parameters is painted in red. {\bf Right:} $\log(\xi/\Lambda)$ as a function of $m_\lambda$ in the region of masses around 10~MeV. The red, black-dashed, blue-dotted, green-long-dashed curves represent $Y/Y_0=1.00,~0.75,~0.50,~0.25$, respectively.}
    \label{Sensitivity_Relic_Coupling}
\end{figure}

With the $\chi^2$ function presented in the Sec. \ref{Simulation_Methodology}, we compute the sensitivity regions displayed in Fig. \ref{Sensitivity_Relic_Coupling}. 
The plot on the left shows in red the space of parameters allowed at 90\%~CL. It covers a wide range of masses, from $10^{-26}$~eV to $10^{12}$~eV. This range includes masses investigated in other contexts~\cite{Pereira:2018xyl,colet}. Heavier masses constrain the interaction to smaller values of $\xi/\Lambda$. The plot on the right shows a smaller range of masses in order to evince the $Y$ sensitivity.
The red curve represents points for which $Y/Y_0 = 1$, black-dashed for $Y/Y_0 = 0.75$, blue-dotted for $Y/Y_0 = 0.5$, and the green-long-dashed for $Y/Y_0 = 0.25$. As expected, a smaller elko-dark matter ratio is related to less sensitivity, i.e., to a looser interaction constraint.

The results show that the most restrictive configuration of parameters are those that concomitantly have a higher particle density and bigger elko mass, implying in smaller coupling strength values for a given $\Lambda$, which means weaker neutrino-elko interaction.
It is important to note that neutrino-dark matter interaction is constrained in the cosmological context because of the structure formation in the Universe. Strong couplings would have interfered in the early Universe changing the structure we observe today. However, investigating neutrino-dark matter interaction at a terrestrial particle accelerator, particularly elko as a dark matter component, gives an independent way of constraining the proposed interaction.

\section{Conclusions}

In order to conduct high precision measurements of the neutrino oscillation phenomenon and determine accurately the parameters of the three neutrino oscillation model, long baseline neutrino experiments, such as DUNE, need a very high intensity beam and a near detector. These two ingredients combined give us the capability to conduct high statistics analyses with neutrinos, opening a very welcome opportunity to investigate physics beyond the standard model of particle physics. Even though long baseline experiments aim to test mainly the oscillation physics of neutrinos, we can take advantage of 
the high intensity flux of the long baseline neutrino beam, investigating non oscillation phenomenon. In this case, we can search for signals that are impossible to detect in experiments with a smaller number of neutrino events.

We show that an experiment with the characteristics of the DUNE is able to discover mass dimension one fermionic particle, so called elko, or to put restrictions in the parameter space dictated by the ratio of the coupling strength with the theory energy scale, $\xi/\Lambda$, the elko and dark matter relic density ratio, $Y/Y_0$, and the elko mass, $m_{\lambda}$. 
Particularly, we investigated constraints for a viable elko mass range between $\mathcal{O}(10^{-26})$~eV and $\mathcal{O}(10^{12})$~eV, which is compatible with several dark matter candidates.
%
In the analysis, we conducted a case study using a DUNE-like detector capability to detect elko via the interaction $\nu_\mu + \lambda \to \lambda + \nu_\mu$ which would affect the neutrino propagation from the source to the far detector. This should encourage future search at neutrino experiments for neutrino-elko interactions.

\section*{Acknowledgments}
CAM acknowledges support from FAPESP Grant Process No. 2014\-/\-1\-9\-1\-6\-4-6. JMHS thanks to CNPq grant No. 303561/2018-1 for partial support. LD thanks "Coordenação de Aperfeiçoamento de Pessoal de Nível Superior - Brasil” (CAPES - Finance Code 001) for the financial support.

\end{document}